\documentclass[amsmath,amssymb,pre,twocolumn]{revtex4-1}

\usepackage{graphicx}

\begin{document}

\title{The free and freer \textit{XY} models}

\author{Petter Holme}
 \email{holme@cns.pi.titech.ac.jp}
\affiliation{Tokyo Tech World Research Hub Initiative (WRHI), Institute of Innovative Research, Tokyo Institute of Technology, Nagatsuta-cho 4259, Midori-ku, Yokohama, Kanagawa, 226-8503, Japan}

\author{Y\'erali Gandica}
\affiliation{Laboratoire de Physique Th\'eorique et Mod\'elisation (LPTM), Dep.\ Physics, Universit\'e de Cergy-Pontoise. 95000. Paris, France}

\begin{abstract}
We study two versions of the \textit{XY} model where the spins but also the interaction topology is allowed to change. In the free \textit{XY} model, the number of links is fixed, but their positions in the network are not. We also study a more relaxed version where even the number of links is allowed to vary, we call it the freer \textit{XY} model. When the interaction networks are dense enough, both models have phase transitions visible both in spin configurations and the network structure. The low-temperature phase in the free \textit{XY} model, is characterized by tightly connected clusters of spins pointing in the same direction, and isolated spins disconnected from the rest. For the freer \textit{XY} model the low-temperature phase is almost completely connected. In both models, exponents describing the magnetic ordering are mostly consistent with values of the mean-field theory of the standard \textit{XY} model.
\end{abstract}

\maketitle

\section{Introduction}

The \textit{XY} model is one of the fundamental spin models of statistical physics. Let $G=(V,E)$ be a graph of $N$ vertices $V$ and $M$ edges $E$ (connected pairs of vertices); let every node $i$ be associated with a spin variable $\Theta=\{\theta_i\}_{i\in V}$, $\theta_i\in [0,2\pi)$. The \textit{XY} model on this graph is defined via the Hamiltonian
\begin{equation}\label{eq:hamiltonian}
    H(G,\Theta)= -J\sum_{(i,j)\in E} \cos (\theta_i - \theta_j) ,
\end{equation}
where $J$ is a coupling constant (that could be $N$-dependent),
and the Boltzmann distribution saying that the probability of the spin configuration $\Theta$ at temperature $T$ is
\begin{equation}\label{eq:boltzmann}
\frac{\exp(-H/T)}{Z} ,
\end{equation}
where the normalization constant $Z$ is called the \textit{partition function}. In our analysis, we put the coupling strength, the Boltzmann's constant, and all the constants of a physical dimension to one, as customary in theoretical studies.

Traditionally, the \textit{XY} model has primarily been used as a model for superconductors and superfluids~\cite{minnhagen1987rmp}. Most famously Kosterlitz, Thouless~\cite{kt:trans} and, independently, Berezinskii~\cite{berezinskii} found that---although there cannot be any regular type of spin ordering in less than three dimensions~\cite{mermin_wagner}---there can be a topological type of order where the spins form vortices that are bound in pairs. Kosterlitz and Thouless were awarded the 2016 Nobel Prize for this discovery. The \textit{XY} model has also been used to model (for a physicist) exotic systems such as birdflocks~\cite{xybird} and discrete-event simulations~\cite{rikvold}.

In classical statistical physics, the underlying topology of the \textit{XY} model is a regular lattice. However, when Networks Science became popular among statistical physicists around the turn of the millennium~\cite{barabasi:book,mejn:book}, there were many studies of spin models~\cite{dorogovtsev2008critical}, including the \textit{XY} model, on various network topologies ~\cite{our:xy,our:dynxy}, contributing with new insights on topology-dependence to the literature of phase transitions. Some early studies, for example, concluded that if the pathlengths of the underlying networks exhibits a logarithmic scaling (as opposed to the geometric scaling of lattices), then the \textit{XY} model, like high-dimensional lattices, shows a mean-field behavior. More recently, De Nigris and Leoncini found that if one allows the number of links to scale non-linearly with $N$, then one can find phases not observed in lattice models~\cite{de2013critical,Nigris_2013}. Expert \textit{et al.}~\cite{expert} mapped the \textit{XY} model to a dynamic model, whose time series, they argued, could characterize the phases of the \textit{XY} model on arbitrary networks. The paper most akin to the current work, however, is Ref.~\cite{our:yx} that investigates the \textit{YX} model, as they call it. The model where the links of the graph, rather than the spins, are updated. Yet other extensions of the \textit{YX} model includes studying it on a geometry with negative curvature~\cite{hepta}, or extending the dimensionality of spins~\cite{stanley}.

In this paper, we study the equilibrium \textit{XY} model where neither the spins, nor the links between them, are fixed. Almost everything is allowed to vary. The only constraints that we impose are that the number of nodes ($N$) and links ($M$) are fixed, and that $G$ should be simple (i.e., there should be no multiple links or self-links). We call this model the \textit{free XY model}. Additionally, we also study the model where $M$ is allowed to vary. Naturally, we call this the \textit{freer XY model}. The coupling constant is rescaled by N, for the freer \textit{XY} model, in order to ensure the energy is an extensive quantity, by analogy to other fully connected (at least in the ground state) spin models like the Sherrington-Kirkpatrick spin glass~\cite{panchenko,our:xy}. An optional definition of $J$ would be $J=N/M$, but as we will see below, $M$ is proportional to $N(N-1)/2$ for all temperatures so this choice would not matter beyond the value of the critical temperature.

\begin{figure*}
\includegraphics[width=0.75\linewidth]{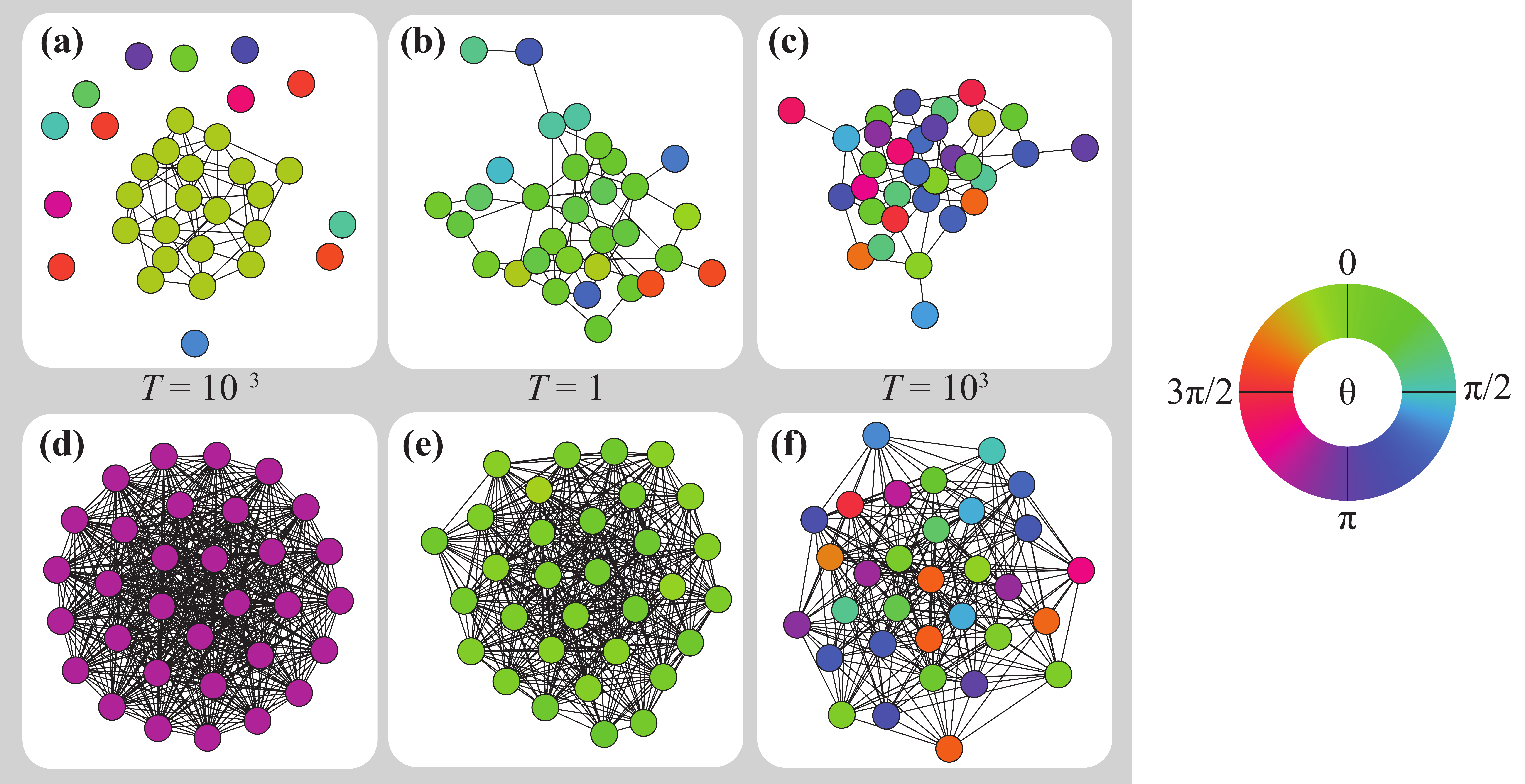}
\caption{Low (a and d), middle (b and e) and high (c and f) temperature configurations of the free \textit{XY} model (a, b, c) with $N=32$ and $M=64$ and the freer \textit{XY} model (d, e, f) with $N=32$. The colors represent the spin angles.}\label{fig:ex}
\end{figure*}

\section{Preliminaries}

\subsection{Monte Carlo simulations}

In this section we will go through details of the Monte Carlo simulations. It is quite technical and could be skipped for a reader who is only interested in the model (not how to sample it). The code we use is available at \url{github.com/pholme/freexy}. At that page we discuss technicalities of the implementation such as how we initialize the configurations (before thermalization).

To simulate the free and freer \textit{XY} model we apply the standard Metropolis rule to the following three update steps:
\begin{enumerate}
    \item For node $i$, replace $\theta_i$ by a new trial angle $\theta_i$'. We let $\theta_i=\theta + \delta\theta$, where $\delta\theta$ is a uniformly distributed random number in the interval $[-\Delta\theta,\Delta\theta]$. We use a Fermi function (see the code for details) to determine $\Delta\theta$ (this is to get interments acceptance rates---if $\Delta\theta$ is too large for low temperatures, the acceptance rate is too small, and vice versa for high temperatures).
    \item For an edge $(i,j)$, find an unconnected node pair $(i',j')$ and replace $(i,j)$, in $E$, by $(i',j')$.
    \item (Only for the freer \textit{XY} model.) For a node pair $(i,j)$, chose a next state as $(i,j)\in E$ or $(i,j)\not\in E$ with equal probability.
\end{enumerate}
Step 1 and 2 ensures ergodicity for the free \textit{XY} model; step 1 and 3 ensures ergodicity for the freer \textit{XY} model. For the freer \textit{XY} model, we keep step 2 (without it, the acceptance rates would be too low at low temperatures). Step 1 and 3 also ensures the system is as random as possible (i.e.\ the random choice in step 3) at the highest temperatures (highest acceptance ratios), this will correctly sample the free and freer \textit{XY} models for reasonably short number of updates. (Otherwise, an easy mistake would be to first try to add, then try to delete, links---at very high temperatures, the second step would just undo the first.) One Monte Carlo \textit{sweep} in our simulations consist of running step 1 for all nodes 10 times, then step 2 for all node edges and (for the freer \textit{XY} model) step 3 for all node pairs. The reason that the spin updates are performed more often is that the small trial angles at low temperatures make the simulation dynamics too slow otherwise.

The energy landscape of the free and freer \textit{XY} models is probably not so complex. Still, we use \textit{exchange Monte Carlo}~\cite{xmc} to manage the temperatures. This is a kind of parallel tempering scheme, where $n_T$ replicas of the system are simulated at $n_T$ different temperatures in parallel. At some different times replicas at adjacent temperatures are swapped. Let
\begin{equation}\label{xmc}
    \Delta = \left(\frac{1}{T}-\frac{1}{T'}\right)(E'-E),
\end{equation}
then, the probability for a swap is given by the Metropolis-like condition
\begin{equation}\label{metro}
\left\{\begin{array}{ll}
1 & \mbox{if $\Delta<0$}\\
\exp(-\Delta) & \mbox{otherwise} \end{array}\right. .
\end{equation}
Like other parallel tempering schemes, this guarantees the system not to be stuck in local energy minima since the replicas perform a random walk in the temperature space. One advantage with exchange Monte Carlo is that (after the initial thermalization), it will always follow the Boltzmann distribution, even immediately after the temperature swaps.

Another advantage of exchange Monte Carlo is that one can use the random walk feature to define a independence criterion. When every replica has visited one quarter (or half, for the first thermalization updates), it is sufficiently updated to be called independent of the previous save. We measure quantities after every sweep, but save the averages of these quantities after the above independence criterion is fulfilled. Finally, we use 1000 independent averages to calculate pooled averages and standard errors that we use in our analysis. Since larger networks have larger energy gaps, they will change temperatures more rarely. This means that the saves happen less frequently for larger networks, which makes sense, because there could be non-local effects that hinders the updates to propagate through the system. It also means, we cannot sample very large networks. One could probably relax the independence criterion (now every spin and every node pair is updated several thousand times between each save), but then it is hard to say whether the samples are statistically independent or not.

After every Monte Carlo sweep, we measure a number of quantities describing the system. When every replica has traversed more than $1/4$ of the temperature levels ($1/2$ the first time, to ensure thermalization), we save averages quantities. We confirm that these averages do not display any autocorrelations (indicative of too frequent measurements versus Monte Carlo updates). These averages---at least 1000 of them for every data point (except the freer \textit{XY} model for $N=512$ where we only use $150$ averages)---are then the basis for the statistics that we present. 

\begin{figure}
\includegraphics[width=\linewidth]{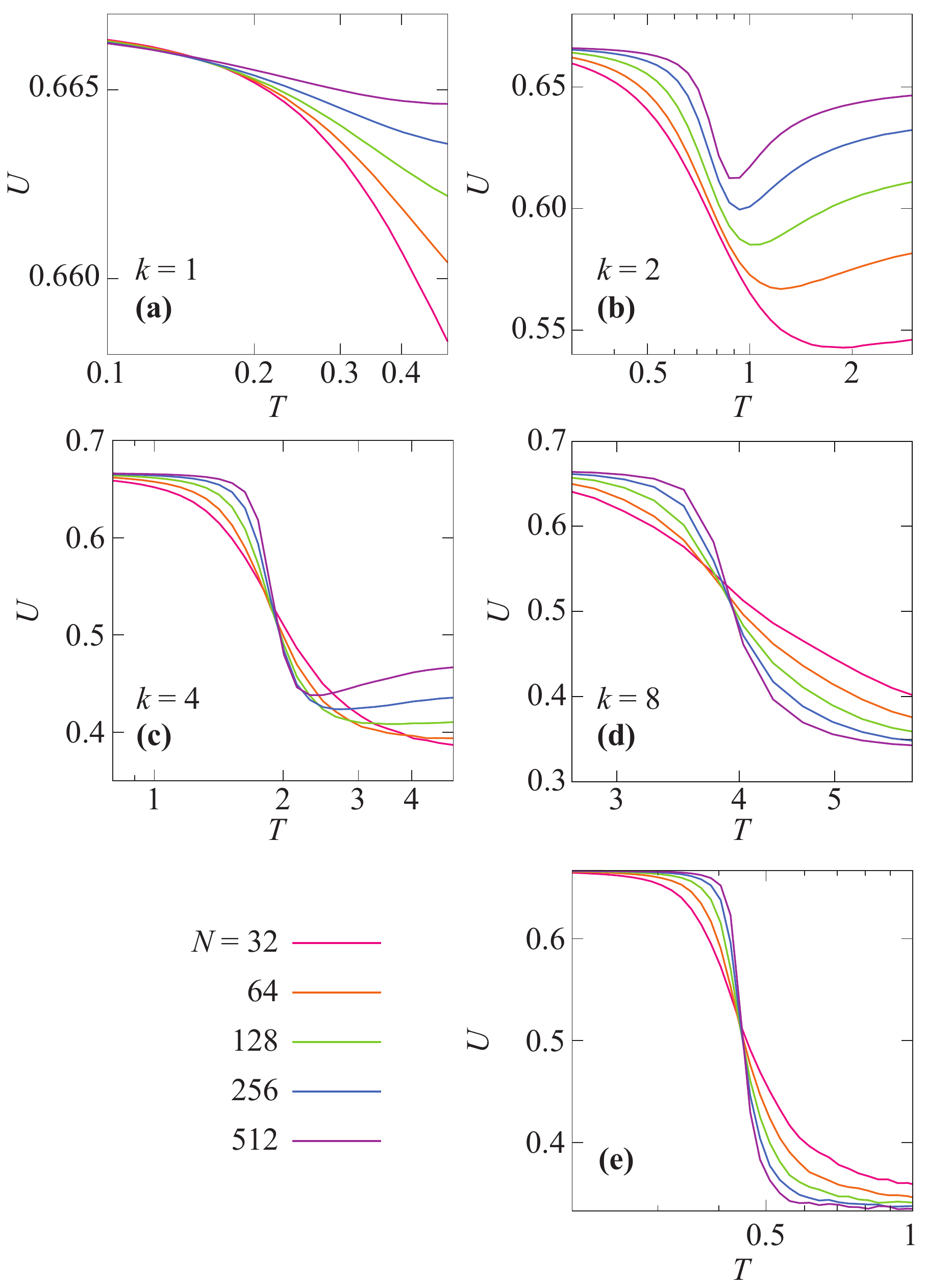}
\caption{Binder's cumulant for the free [panel (a)--(d)] and freer [panel (e)] \textit{XY} models as a function of temperature for various system sizes. For the free \textit{XY} model, we plot four values of the average degree---$k=1$ in panel (a), $k=2$ in panel (b), $k=4$ in panel (c), and $k=8$ panel (d). The $T$-axes are logarithmic.}\label{fig:binder}
\end{figure}

\begin{figure}
\includegraphics[width=\linewidth]{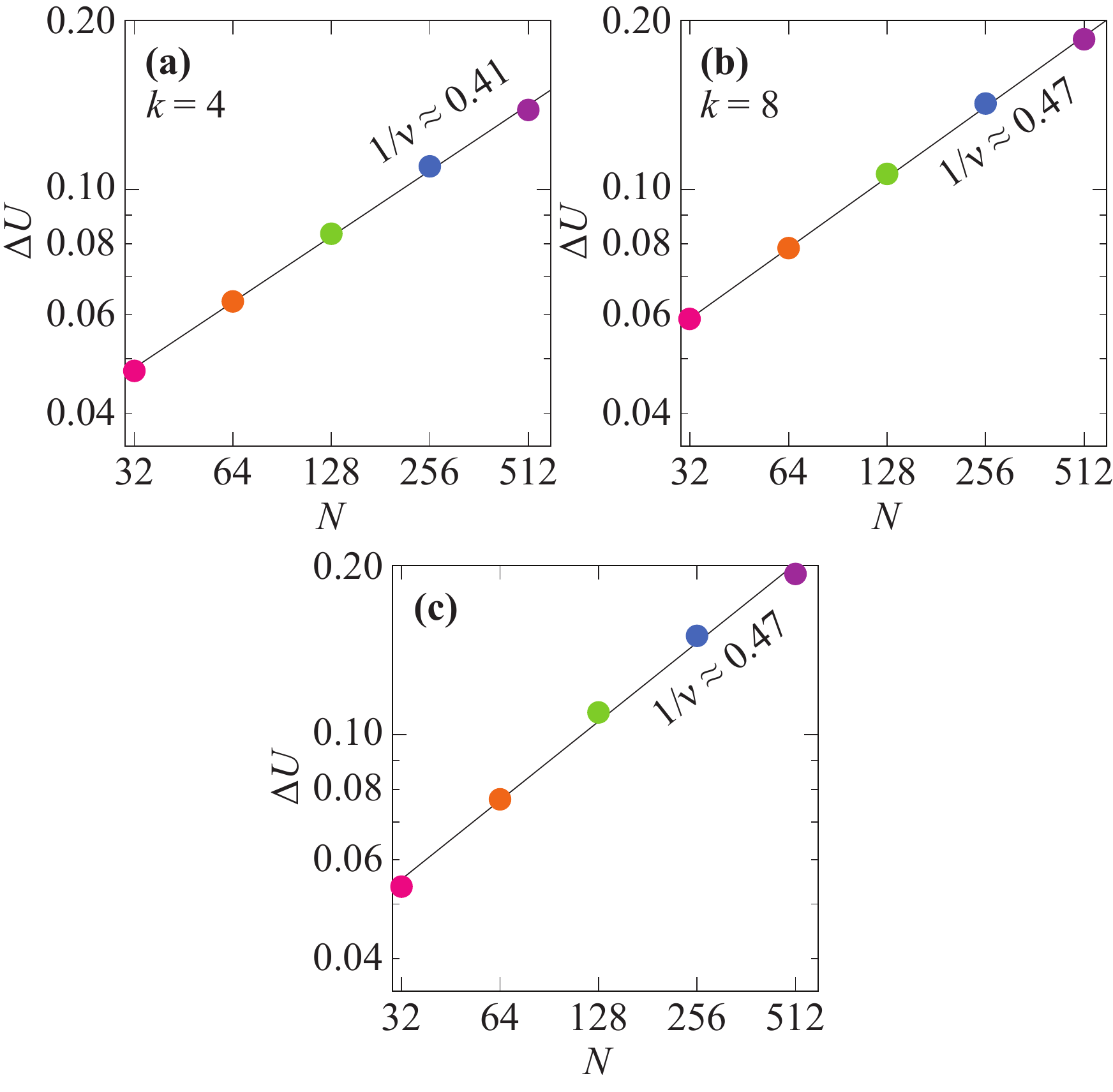}
\caption{Determining the critical exponent $\nu$ by finite size effects on $U$ near the critical point. For the free \textit{XY} model [panels (a) and (b)---the former for $k=4$, the latter for $k=8$] and the freer \textit{XY} model [panel (c)]. Colors define the same system size as in Fig.~\ref{fig:binder}. The axes are logarithmic. }\label{fig:binder_expo}
\end{figure}

\begin{figure}
\includegraphics[width=\linewidth]{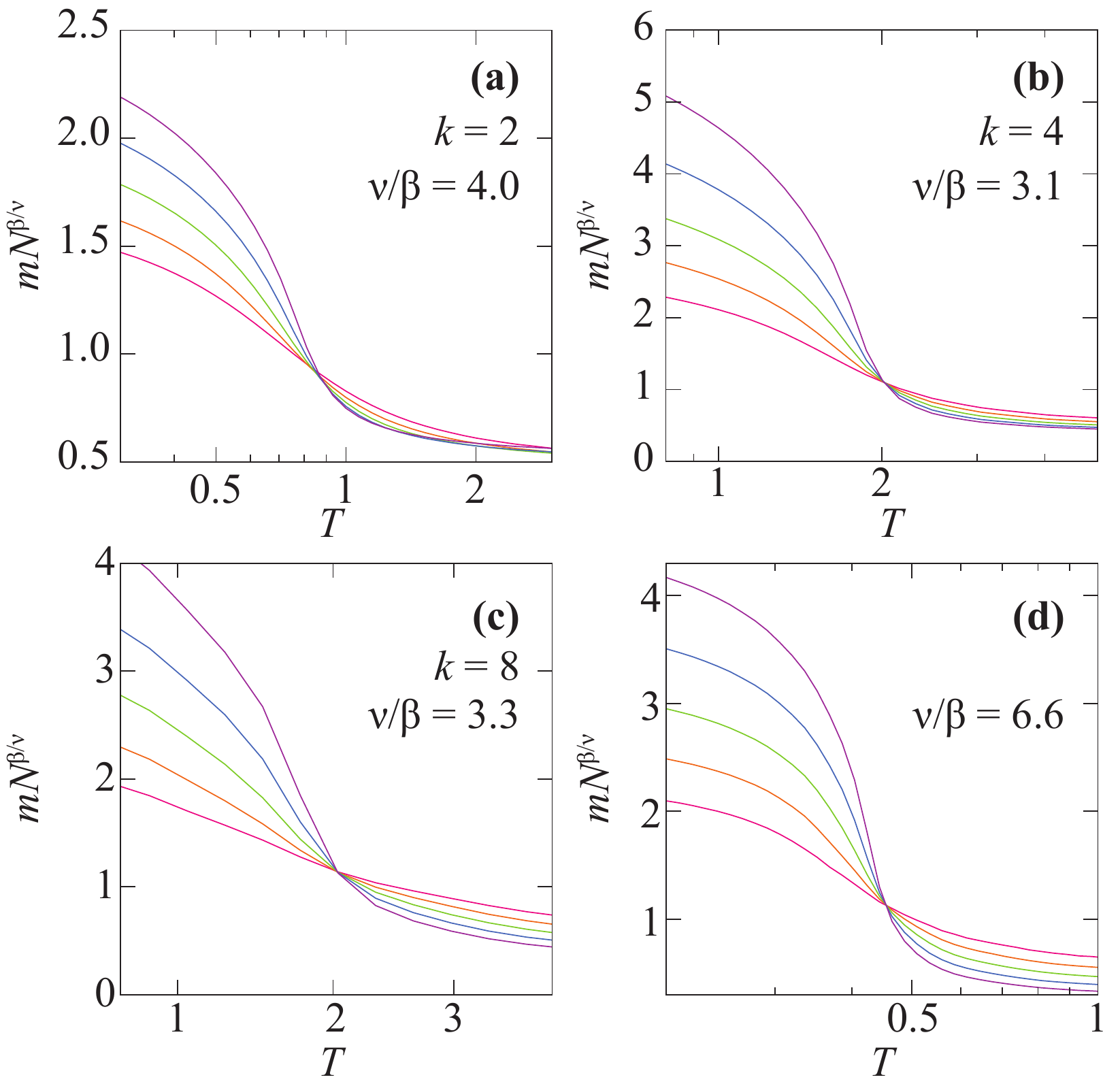}
\caption{Crossing plots for the magnetization to determine $T_c$ and the exponent $\nu$. For the free \textit{XY} model [panels (a)--(c)] and the freer \textit{XY} model [panel (d)]. The lines have the same color as in Fig.~\ref{fig:binder}. The $T$-axes are logarithmic.}\label{fig:mag_cross}
\end{figure}

\subsection{Quantities}

After every Monte Carlo sweep, we measure a number of quantities describing the system. In order to have the complete scenery we present, in this section, results for the -structural-topological- equilibrium configurations along with the regular functions describing the spins-alignment. 

\subsubsection{Network structure}

From inspecting the networks (Fig.~\ref{fig:ex}(a-c)), is visible how the free \textit{XY} passes from a magnetized phase, characterized by one system-size dense cluster with similar spins and few isolated disoriented nodes, at low temperature to a disorder phase, constituted by random angles-nodes on  Erd\H{o}s-R\'enyi-like networks, at high temperatures. In the freer \textit{XY} model, the number of links goes from the maximum ($N(N-1)/2$) for low temperatures to ($N(N-1)/4$) for high temperatures, and the networks are very rarely fragmented.

Based on the above observations, we measure only the simplest metrics to characterize the networks structures: The number of components $c$, the average size of the largest $S$ and second largest components $S_2$, the diameter (the longest shortest path within a connected component---which in random enough networks carries the same information as the average distance~\cite{fan_chung}). We add to our analysis, the number of links for the freer XY model. 

\subsubsection{Characterizing the spin configuration}

The magnetization, $m$, is the fundamental order parameter for the \textit{XY} model on lattices in three and more dimensions. For a connected network, the definition is
\begin{equation}\label{eq:simple_mag}
    m = \left| \left\langle e^{i\theta_i}\right\rangle \right|.
\end{equation}
the average is over all nodes in the network. Our networks could be split into mutually disconnected components---$G=G_1,\dots,G_c$. The most straightforward generalization of Eq.~\eqref{eq:simple_mag} is an average weighted by the number of nodes of the components ($N_i$ for component $i$):
\begin{equation}\label{eq:mag}
m = \frac{\sum_c N_c m_c}{N}  ,
\end{equation}
where $N_c$ is the number of nodes, and $m_c$ is the magnetization of component $c\subset G$.

To perform the finite-size scaling analysis on our system, we start the fluctuations-based study by the commonly used, Binder's cumulant, 
\begin{equation}\label{eq:binder}
U=1-\frac{\langle m^4\rangle}{3\langle m^2\rangle^2},
\end{equation}
measuring the kurtosis of the magnetization distribution. The averages are taken over difference samples of configurations in the Monte Carlo simulations.

Another quantity characterizing the fluctuations of the magnetization is the \textit{susceptibility}
\begin{equation}
    \chi=\frac{\langle m^2\rangle-\langle m\rangle^2}{T}.
\end{equation}
We also measure the \textit{specific heat} which is related to the fluctuations in the energy:
\begin{equation}
    C_V=\frac{\langle H^2\rangle-\langle H\rangle^2}{NT^2}.
\end{equation}
Both the specific heat and the susceptibility diverges at most phase transitions (notably not at the Berezinskii-Kosterlitz-Thouless transition of the two-dimensional \textit{XY} model~\cite{minnhagen1987rmp}).

\begin{figure}
\includegraphics[width=\linewidth]{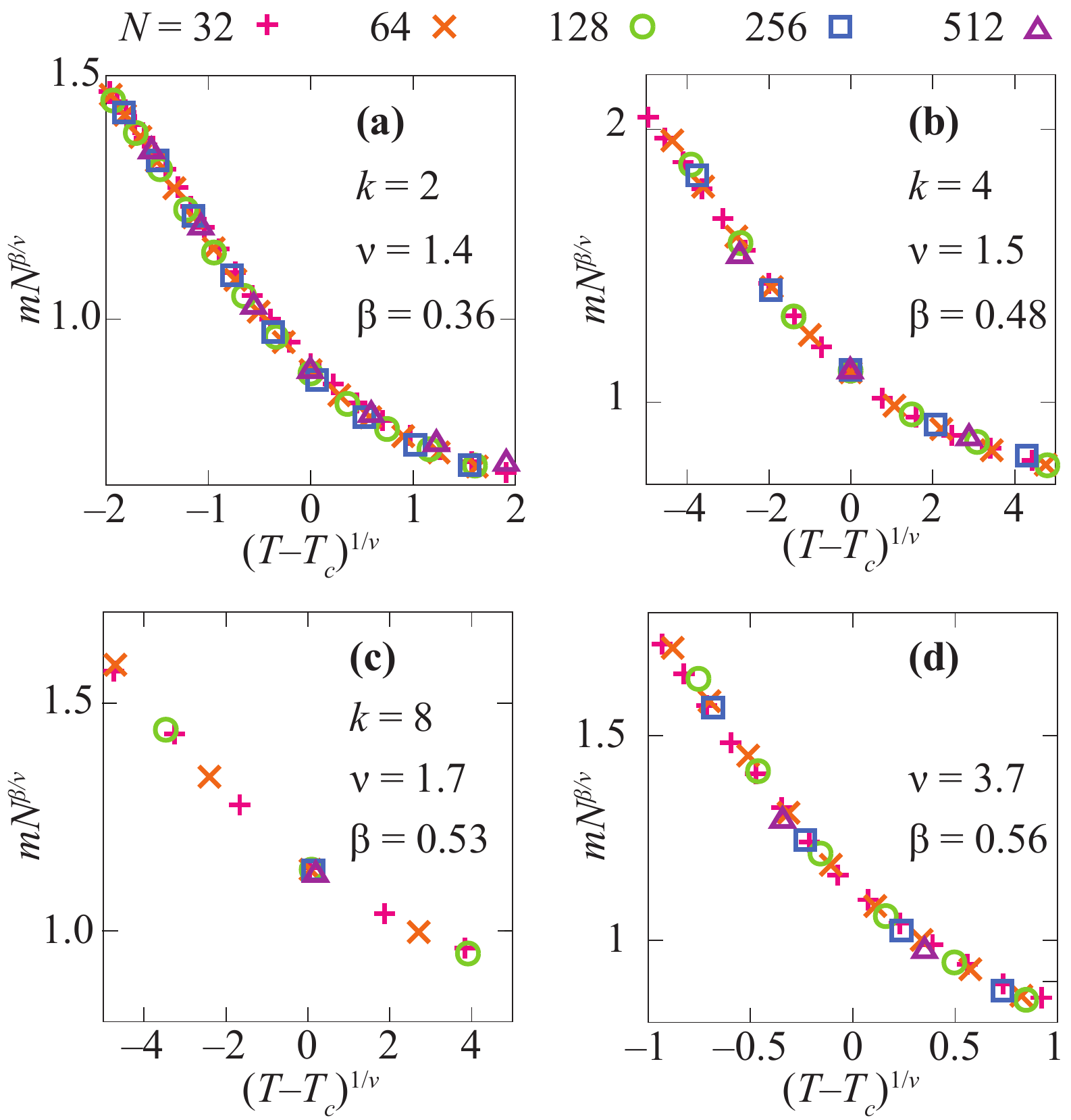}
\caption{Collapse plots to determine $\nu$ and $\beta$. For the free \textit{XY} model [panels (a) and (b)---the former for $k=4$, the latter for $k=8$] and the freer \textit{XY} model [panel (c)]. }\label{fig:mag_collapse}
\end{figure}

\begin{figure}
\includegraphics[width=\linewidth]{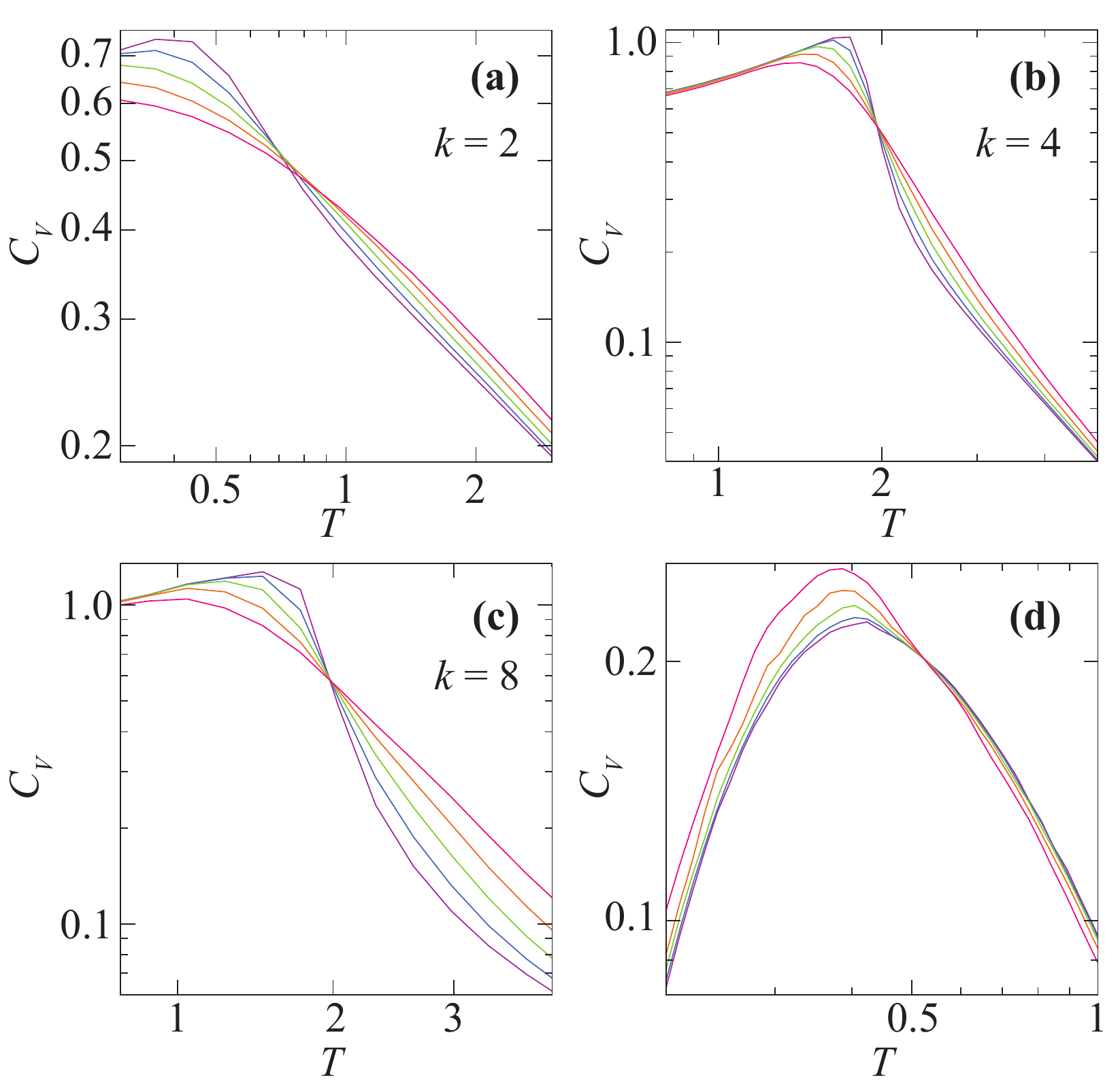}
\caption{The specific heat for the free [panel (a)--(d)] and freer [panel (e)] \textit{XY} models as a function of temperature for various system sizes. For the free \textit{XY} model, we plot the results for $k=2$ in panel (a), $k=4$ in panel (b), $k=8$ in panel (c), and the freer \textit{XY} model panel (d). The lines symbolize the same as in Fig.~\ref{fig:binder}. Note the axes are logarithmic.}\label{fig:spec_heat}
\end{figure}

\begin{figure}
\includegraphics[width=\linewidth]{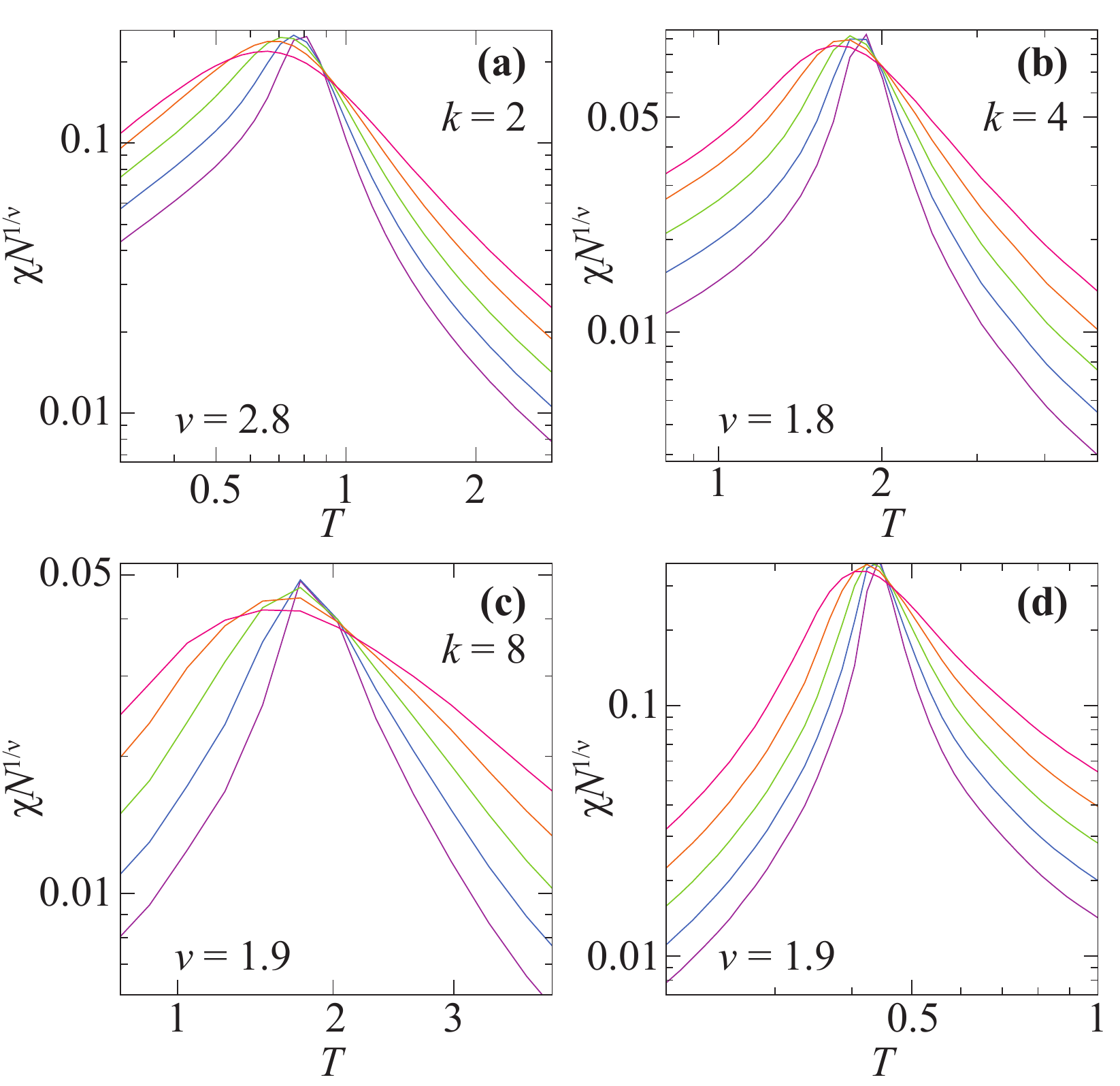}
\caption{The crossing plot for the susceptibility of the free [panel (a)--(d)] and freer [panel (e)] \textit{XY} models as a function of temperature for various system sizes. For the free \textit{XY} model, we plot the results for $k=1$ in panel (a), $k=2$ in panel (b), $k=4$ in panel (c), and $k=8$ panel (d). The lines symbolize the same as in Fig.~\ref{fig:binder}. The axes are logarithmic.}\label{fig:susc_cross}
\end{figure}

\section{Results}

In this section we present our simulation results. Most plots are very well-converged, and error bars would be smaller than the line width or symbol size. We omit them throughout the analysis, but comment on the case with larger errors.

\subsection{Ferromagnetic transition}

We start by showing the thermodynamic functions to analyze the magnetic transition of the \textit{XY} model (not quantities relating to the BKT transition, since there can be no notion of vortices in networks). We will stick to the notation of the literature of the \textit{XY} model. This means that some exponents that are known to be same---such as $\nu$ that describes the critical behavior both of Binder's cumulant and the susceptibility---could actually be different.

\subsubsection{Binder's cumulant}

At a ferromagnetic phase transition, the value of Binder's cumulant will be independent of $N$. Therefore, it is a convenient quantity for determining the critical temperature. In Fig.~\ref{fig:binder}, we plot $U$ for both the free and freer \textit{XY} models. For the freer \textit{XY} model, the denser the networks ($k=4$ of panel (b) and $k=8$ panel (c)), the clearer the crossing point of the $U(T)$ curves. For $k=4$ this happens at $T_c= 1.93\pm 0.03$, while at $T_c= 3.91\pm 0.03$ for $k=8$. These values are also close to the mean-field values---for a lattice where every spin has $k$ neighbors, the mean-field transition happens at temperature $T_{\rm MF}=k/2$, so $T_c = 0.96 T_{\rm MF}$ and $T_c = 0.98 T_{\rm MF}$ respectively. Probably $T_c/T_{\rm MF}$ approaches one when increasing $k$.

This supports the conclusion that the free \textit{XY} model undergoes a phase transition primarily akin to the mean-field ferromagnetic transition of the \textit{XY} model on high-dimensional lattices. This conclusion is corroborated by the observation that the critical exponent $\nu$ (related to the divergence of the correlation volume) is close to the mean-field value $\nu_{\rm MF}=2$~\cite{our:xy}, see Fig.~\ref{fig:binder_expo}. We use the relationship
\begin{equation}
\Delta U \approx \text{constant} \times N^{1/\nu} ,
\end{equation}
where $\Delta U = U(T_c+\delta T) - U(T_c-\delta T)$ for small $\delta T$. We observe $\nu=2.4\pm 0.1$ for $k=4$ and $\nu=2.1\pm 0.1$ for $k=8$, which are slightly larger than, but still compatible with the mean-field value $\nu_{\rm MF}=2$~\cite{our:xy}.

For the sparser networks of $k=1$ and $k=2$, Binder's cumulant does not have a crossing point. For these values of $k$, the networks are rather fragmented. Recall that Erd\H{o}s-R\'enyi random graphs have their transitions between being fragmented and having a giant component (a largest connected component scaling like $N$) at $k=2$. At $k=1$, the network is so sparse that the largest possible connected component is $N/2$. In practice, however, it is much smaller (see Fig.~\ref{fig:size_lcc}), so effectively there can be no long-range spin correlations, and there is thus no wonder that the behavior of the model at such parameter values is very far from the large-$N$ values.

For the freer \textit{XY} model, there is also a clear crossing of the Binder's cumulant; see Fig.~\ref{fig:binder}(e). Also in this case, the critical temperature $T_c = 0.446 \pm 0.005$ is slightly lower than the mean field value $T_{\rm MF}=1/2$ (recall that the freer \textit{XY} model's Hamiltonian is divided by $N$, if we used the same form for the free \textit{XY} model, $T_{\rm MF}$ would be $1/2$ for that as well).

\subsubsection{Magnetization}

The order parameter---the magnetization---, is another way to analyse the critical behavior on the phase transition, which close to $T_c$ is expected to follow the scaling relation
\begin{equation}\label{eq:mag_fss}
m=N^{-\beta/\nu}f\left(T N^{1/\nu}\right),
\end{equation}
for a smooth function $f$. This means that the correct choice of $\nu$ would make curves of $mN^{\beta/\nu}$ cross at $T_c$. As seen in Fig.~\ref{fig:mag_cross}, this is possible for $k=4$ and $k=8$, where Binder's cumulant showed a crossing of the curves. The obtained critical temperatures $T_c=2.01\pm 0.05$ (for $k=4$) $T_c=4.03\pm 0.06$ (for $k=8$) are also consistent with the predictions from mean-field theory. We find the scaling exponent $\nu/\beta$ to be $\nu/\beta=3.1\pm 0.3$ ($k=4$) $\nu/\beta=3.3\pm 0.3$ ($k=8$). For the freer \textit{XY} model we get $T_c=0.453\pm0.002$ with $\nu/\beta=4.0\pm 0.2$. 

Unlike the analysis of Binder's cumulant, we can obtain a crossing plot for $k=2$ as well (but none for $k=1$). See Fig.~\ref{fig:mag_cross}(a). This happens at a temperature further from the mean-field prediction: $T_c=0.87\pm0.01$ and at a much larger value of the exponent $\nu=6.6\pm0.1$.

From Eq.~\eqref{eq:mag_fss}, also follows that we can determine $\beta$ by finding a value that collapses all curves to one. Our estimates of $\beta$---giving the collapse plots seen in Fig.~\ref{fig:mag_collapse}. The values are: $\beta=0.36\pm 0.05$ ($k=2$), $\beta=0.48\pm 0.02$ ($k=4$) and $\beta=0.53\pm 0.02$ ($k=8$). Once again, for the denser networks these values are close to the mean-field value $\beta_{\rm MF}=1/2$~\cite{our:xy}. The freer \textit{XY} model has the beta value $\beta=0.56\pm 0.04$.

The values of $\nu$ are smaller than the mean-field value of $\nu_{\rm MF}=2$. We will not dwell much on this discrepancy more than noting two scenarios: Either it will disappear for yet larger system sizes; or (more likely) the extra flexibility from the underlying dynamic network changes the universality class of the model. Note that there is no \textit{a priori} reason that the exponents 

\subsubsection{Specific heat}

A third way of monitoring phase transitions is by the specific heat. This is another quantity that is scale-independent at $T_c$ for mean-field transitions. Like for the magnetization, for $k=2$ there is a crossing (although somewhat blurry) at $T_c=0.95\pm0.05$. See Fig.~\ref{fig:spec_heat}. For $k=4$ and $k=8$, the crossing is very sharp at $T_c=1.97\pm0.01$ ($k=4$) and $T_c=3.99\pm0.01$ ($k=8$), respectively. For the freer \textit{XY} model, $C_V$ also gives a crossing at $T_c = 0.52 \pm 0.06$ but the curves follow different functional forms than the free \textit{XY} model, the scaling is also reversed (so that larger system sizes lies above the smaller ones below $T_c$, and above them for $T>T_c$) and the crossing is less clear.

\subsubsection{Susceptibility}

The susceptibility $\chi$ is known to follow a scaling relation
\begin{equation} \label{eq:chi_fss}
    \chi\sim N^{-1/\nu} f(T-T_c) ,
\end{equation}
for temperatures close to $T_c$. In Fig.~\ref{fig:susc_cross}, we do indeed find crossings around the same values of $T_c$ as determined by Binder's cumulant and the magnetization---$T_c=0.91\pm0.04$ for $k=2$; $T_c=1.98\pm0.06$ for $k=4$; $T_c=4.1\pm0.1$ for $k=8$; and $T_c=0.44\pm0.04$ for the freer \textit{XY} model. The values of the exponent $\nu$ that we obtain in this way are closer to the mean-field value of $\nu=2$ than the ones we obtained from the crossing plots for the magnetization---$\nu=1.8\pm0.1$ for $k=2$; $\nu=1.9\pm0.1$ for $k=4$; $\nu=1.9\pm0.1$ for $k=8$; and $\nu=2.1\pm0.2$ for the freer \textit{XY} model. A striking difference between Fig.~\ref{fig:susc_cross} and the mean-field behavior however is that $\chi N^{1/\nu}$ does not seem to diverge (or does so extremely slowly) as the temperature approaches criticality from below.

\begin{figure}
\includegraphics[width=0.8\linewidth]{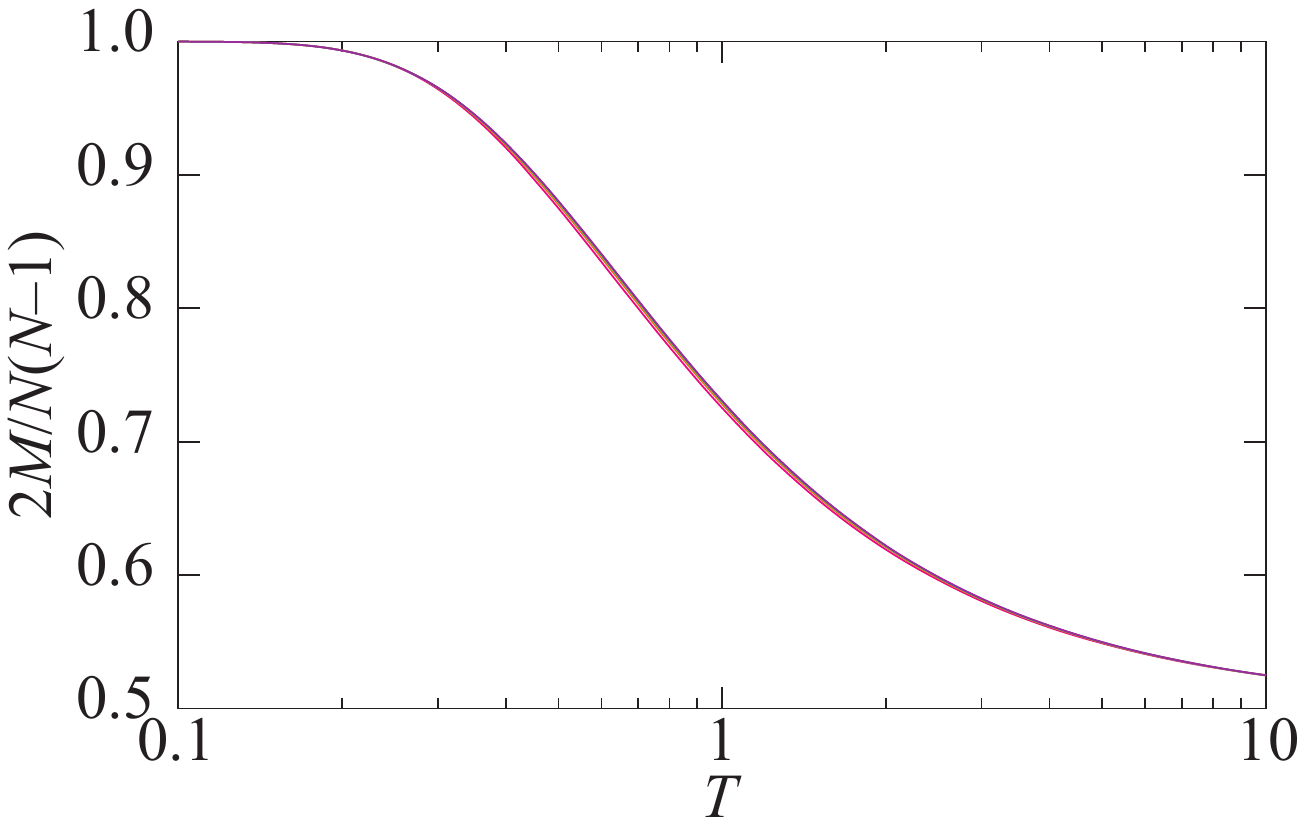}
\caption{The connectance (number of links $M$ divided by the maximal possible number of links $N(N-1)/2$) in the freer \textit{XY} model as a function of temperature. The lines symbolize the same as in Fig.~\ref{fig:binder}. The $T$-axis is logarithmic.}\label{fig:nlinks}
\end{figure}

\begin{figure}
\includegraphics[width=\linewidth]{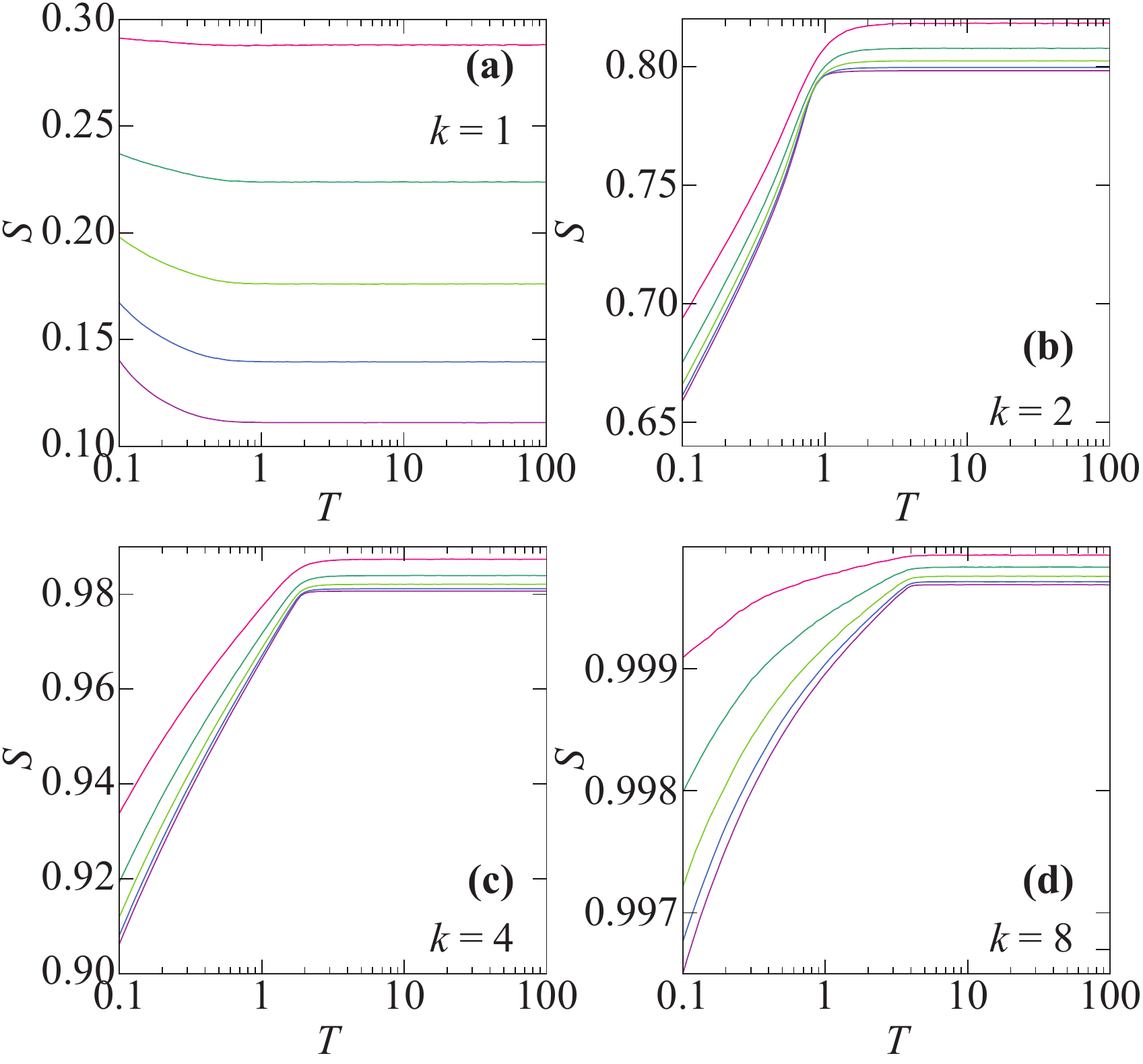}
\caption{The size of the largest connected component of the free \textit{XY} model as a function of temperature for various system sizes. We show results for $k=1$ in panel (a), $k=2$ The lines symbolize the same as in Fig.~\ref{fig:binder}. in panel (b), $k=4$ in panel (c), and $k=8$ panel (d). The $T$-axes are logarithmic.}\label{fig:size_lcc}
\end{figure}

\begin{figure}
\includegraphics[width=\linewidth]{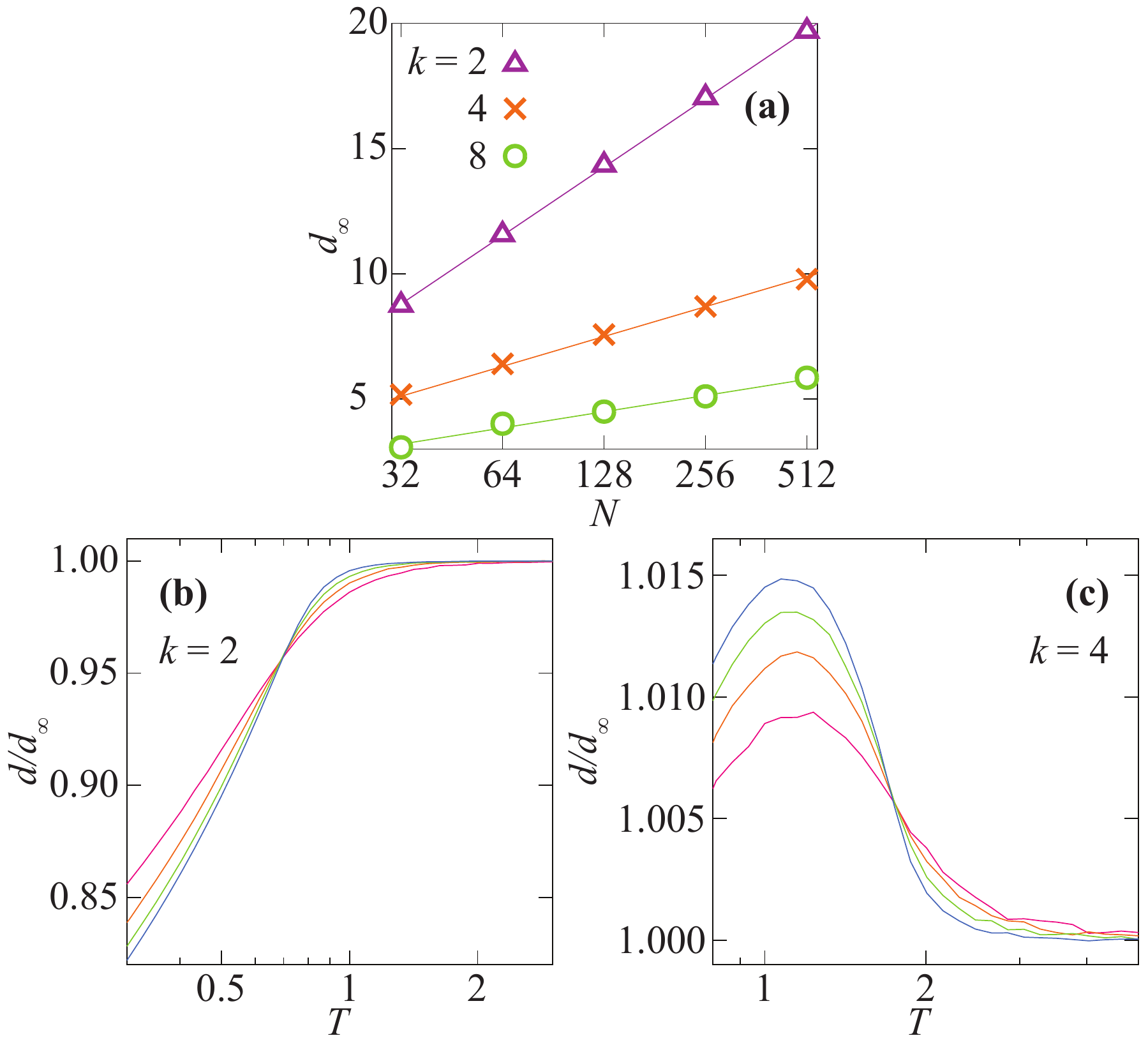}
\caption{Panel (a) shows the logarithmic scaling of the diameter of an Erd\H{o}s-R\'enyi model of the same $N$ and $M$ (i.e.\ the diameter for infinite temperature), $d_\infty$. Panels (b) and (c) shows the diameter relative to $d_\infty$ for $k=2$ (b) and $k=4$ (c). The lines symbolize the same as in Fig.~\ref{fig:binder}. The $T$-axes are logarithmic.}\label{fig:dia}
\end{figure}

\subsection{Network structural effects}

In this section, we search phase transitions in the network structure.

\subsubsection{Number of edges}

This first analysis applies only to the freer \textit{XY} model, where the number of links is allowed to changed. For low temperatures it is clearly favorable to have as many links as possible. For high temperatures, a link exists or not with equal probability. Thus we can see that the connectance (the fraction of node pairs that have a link) goes from $1$ at zero temperatures to $1/2$ at $T=\infty$. This is confirmed in Fig.~\ref{fig:nlinks}. Interestingly, there are almost no size dependence in this quantity---it goes smoothly from one limit to the other. This also means that an alternative freer \textit{XY} model with $J=N/M$ would have a phase transition with the same exponents, but (since the temperature is measured in units of $J$) a different critical temperature.

\subsubsection{Size of the largest component}

Probably the most common quantity to characterize phase transitions in networks is the size of the largest connected component~\cite{mejn:book}. We use the fraction of nodes in the largest component $S$ for our analysis. The freer XY model is so dense (Fig.~\ref{fig:nlinks}) that it is basically only fully connected (only for $N=8$, we were able to observe fragmented networks). Thus, in this section, we focus on the free \textit{XY} model for our further analysis.

In Fig.~\ref{fig:size_lcc}, we plot $S$ for all $k$ values that we investigate. Just like the quantities related to the magnetic order, the $k=1$ case is very different. For $k=2,4,8$ the $S(T)$ curves seems to converge to a plateau at high $T$ accompanied by a monotonic increase for lower temperatures. For $k=1$, however, $S$ just goes to zero for all temperatures as $N$ increases.

When $k\geq 2$, the borders between the different scaling regimes of $S$ seem to be very close the phase transition as indicated by e.g.\ Binder's cumulant. If one just takes the maximum curvature of the $N=512$ curves in Fig.~\ref{fig:size_lcc} as an estimate of $T_c$ one obtains: $T_c=0.9\pm0.1$ for $k=2$, $T_c=1.9\pm0.2$ for $k=4$ and $T_c=3.8\pm0.2$ for $k=8$. These values are consistent with the hypothesis that the free \textit{XY} model has a phase transition visible in both spin-related and network quantities.

\subsubsection{Diameter}

Our final network-structural analysis concerns diameter $d$ of the networks. In the high-temperature limit the free \textit{XY} model networks are effectively Erd\H{o}s-R\'enyi random graphs, thus having a logarithmic scaling of the diameter~\cite{fan_chung}. We use $d_\infty$ to denote the diameter at $T=\infty$. This is shown in Fig.~\ref{fig:dia}(a). To analyze $d$ for lower temperatures, we find it useful to rescale the values by $d_\infty$. For $k=8$, $d/d_\infty$ are so close to one that the noise makes it impossible to identify meaningful trends. The $d/d_\infty$ curves for $k=2$ and $k=4$ are shown in Fig.~\ref{fig:dia}(b) and (c) respectively. Interestingly they both show crossing points (but fot $k=8$ there is not crossing). These crossing points do not, however, coincide with the $T_c$ detected by other quantities. 

This could perhaps point at another structural phase transition, but since there is no such behavior for $k=8$, we believe this be a transient phenomenon that would disappear for larger sizes. Further studies are needed to resolve this issue. Another interesting observation is that the behavior outside of the crossing point is different for $k=2$ and $k=4$. For the former, larger system sizes have smaller $d/d_\infty$ for high temperatures and smaller $d/d_\infty$ for lower temperatures, for the latter, this situation is reversed. How the transition between these situation looks is another interesting open question.

\begin{table}
\caption{\label{tab:tc}Estimates of $T_c$ from different quantities. The last row gives the mean-field values.}
\begin{ruledtabular}
\begin{tabular}{l|llll}
Quantity & $k=2$ & $k=4$ & $k=8$ & \text{freer}\\ \hline
$U$ & -- & 1.93(3) & 3.91(3) & 0.446(5) \\
$m$ & 0.87(1) & 2.01(5) & 4.03(6) & 0.453(2)\\
$C_V$ & 0.95(5) & 1.97(1) & 3.99(1) & 0.52(6)\\
$\chi$ & 0.91(4) & 1.98(6) & 4.1(1) & 0.44(4)\\
$S$ & 0.90(10) & 1.9(2) & 3.8(2) & --\\\hline
MF & 1 & 2 & 4 & $1/2$\\
\end{tabular}
\end{ruledtabular}
\end{table}

\begin{table}
\caption{\label{tab:nu}Estimates of $\nu$ from different quantities. The mean-field value is $\nu_{\rm MF}=2$.}
\begin{ruledtabular}
\begin{tabular}{l|llll}
Quantity & $k=2$ & $k=4$ & $k=8$ & \text{freer}\\ \hline
$U$ & -- & 2.4(3) & 2.1(2) & 2.1(5) \\
$m$ & 1.4(2) & 1.5(2) & 1.7(2) & 3.7(3)\\
$\chi$ & 2.8(3) & 1.8(3) & 1.9(2) & 1.9(3)\\
\end{tabular}
\end{ruledtabular}
\end{table}

\section{Discussion}

We have studied two versions of the \textit{XY} model where the interaction topology is free to change. One (the free \textit{XY} model) where the number of links is fixed; another (the freer \textit{XY} model) where the number of links is also allowed to vary. For the freer \textit{XY} model, and the free \textit{XY} model when it is dense enough, we find a phase transition that is visible both in spin and network related quantities. For low temperatures, the free \textit{XY} model is characterized by tightly connected clusters of spins pointing in the same direction, and isolated spins disconnected from the rest. In the freer \textit{XY} model, in the low-$T$ phase, the system is always connected and the system has close to the maximum number of links. One could imagine that the spin ordering and network fragmentation have transitions at different temperatures, but we find no evidence of that.

To recapture our findings in a bit more detail: for $k>2$ in the free \textit{XY} model and the freer \textit{XY} model, the critical temperatures and exponents obtained from finite-size scaling are close to the values of the mean-field approximation to the standard \textit{XY} model. There are also traces of a transition in the network structural measures. From these observations, we believe there is one transition, primarily driven by spin alignment and similar to in high-dimensional lattices. The network structure could be regarded as following the spin-ordering, rather than the other way around. We get a similar conclusion for critical exponents---the exponent $\nu$ of traditional spin systems~\cite{our:xy} as measured by Binder's cumulant and susceptibility is consistent with the mean-field values, but for magnetization they are consistently lower than those obtained from the magnetization. The one exception to the for the freer \textit{XY} model where $\nu$ is almost twice as large as $\nu_{\rm MF}$.

When $k=2$, the critical temperature, is consistently lower than the mean-field prediction, probably the approximation breaks down at that point but there is still a transition fitting the above description. For even sparser systems (i.e.\ our $k=1$) simulations, we cannot find evidence of a phase transition. There might still be one, or there could be a cross-over behaviour with a continuous change from disorder at high-temperature to fragmented components of aligned spins close to zero temperature. We summarize the measured critical temperatures in Table~\ref{tab:tc} and the $\nu$ values in Table~\ref{tab:nu}.

Perhaps the greatest lesson of this analysis is just how robust the magnetic ordering is in the \textit{XY} model. We know that putting spin models on random networks with their short path-lengths is equivalent to placing them on high-dimensional lattices~\cite{our:xy,hong:comment}. Not even the fairly large perturbation to the original model that our models make manages to change the spin-order transition much from the mean-field one of high-dimensional lattices. The exception to this conclusion is the extreme sparse case we study ($k=1$). Thus, even though the transition is visible in network-structural quantities, it seems to be driven by the spin ordering.

The main open question is what kind of transition the free \textit{XY} model experiences when $k$ decreases. Obviously the system at $k=1$ behaves very differently from the larger $k$ values. One scenario is that there is a phase transition in $k$ (note that $k$ is a continuous parameter); another scenario is that there is a cross-over from the mean-field dominated situation at denser networks.

\acknowledgments{P.H. was supported by JSPS KAKENHI Grant Number JP 18H01655 and by the Grant for Basic Science Research Projects by the Sumitomo Foundation.}

\bibliography{freexy}

\end{document}